\documentclass{esannV2}
\usepackage[dvips]{graphicx}
\usepackage[latin1]{inputenc}
\usepackage{amssymb,amsmath,array}
\usepackage{epsfig}

\newcommand{\bx}{\mathbf{x}}

\newcommand{\bw}{\mathbf{w}}

\newcommand{\bz}{\mathbf{z}}

\newcommand{\bsw}{\boldsymbol{w}}
\newcommand{\bsr}{\boldsymbol{r}}
\newcommand{\bsv}{\boldsymbol{v}}

\newcommand{\bst}{\boldsymbol{t}}

\newcommand{\btheta}{\boldsymbol{\theta}}

\newcommand{\bbeta}{\boldsymbol{\beta}}

\newcommand{\IR}{I\!\!R}
\begin{document}
\title{A regression model with a hidden logistic process for signal parametrization}

\author{F. Chamroukhi$^{1,2}$, A. Sam\'e$^1$, G. Govaert$^2$ and P. Aknin$^1$
%
%
\vspace{.3cm}\\
%
1- French National Institute for Transport and Safety Research (INRETS)\\
2 Rue de la Butte Verte, 93166 Noisy-Le-Grand Cedex (France)
%
%
\vspace{.1cm}\\
2- Compi\`egne University of Technology \\HEUDIASYC Laboratory, UMR CNRS 6599 \\
BP 20529, 60205 Compi\`egne Cedex (France)\\
}

\maketitle

\begin{abstract}
A new approach for signal parametrization, which consists of a specific regression model incorporating a discrete hidden logistic process, is proposed. The model parameters are estimated by the maximum likelihood method performed by a dedicated Expectation Maximization (EM) algorithm. The parameters of the hidden logistic process, in the inner loop of the EM algorithm, are estimated using a multi-class Iterative Reweighted Least-Squares (IRLS) algorithm. An experimental study using simulated and real data reveals good performances of the proposed approach.
\end{abstract}

\section{Introduction}

\begin{figure}[!h]
  \centering
  \includegraphics[height = 4 cm,width=5cm]{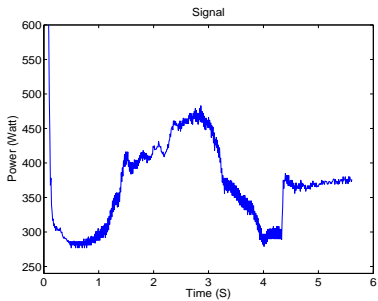}
  \caption{Example of the electrical power consumed during a point operation}\label{signal_intro}
 \end{figure}

In the context of the predictive maintenance of the French railway switches (or points) which enable trains to be guided from one track to another at a railway junction, we have been brought to parameterize switch operations signals representing the electrical power consumed during a point operation (see figure \ref{signal_intro}).
The final objective is to exploit these parameters for the identification of incipient faults.

The method we propose to characterize signals is based on a regression model incorporating a discrete hidden process allowing abrupt or smooth switchings between various regression models. This approach has a connection with the switching regression model introduced by Quandt and Ramsey \cite{quandt} and is very linked to the Mixture of Experts (ME) model introduced by Jordan and Jacobs \cite{jordan HME} by the using of a time-dependant logistic transition function. The ME model, as discussed in \cite {waterhouse}, uses a conditional mixture modeling where the model parameters are estimated by the Expectation Maximization (EM) algorithm \cite{dlr}\cite{mclachlan EM}. Other alternative approaches are based on Hidden Markov Models in a context of regression \cite{fridman}. A dedicated EM algorithm including a multi-class Iterative Reweighted Least-Squares (IRLS) algorithm \cite{irls} is proposed to estimate the model parameters.

Section 2 introduces the proposed model and section 3 describes the parameters estimation via the EM algorithm. The fourth section is devoted to the experimental study using simulated data and real data.

\section{Regression model with a hidden logistic process}
\label{sec: regression model}

\subsection{The global regression model}
We represent a signal by the random sequence $\bx = (x_1,...,x_n)$ of $n$ real observations, where $x_i$ is observed at time $t_i$.
This sample is assumed to be generated by the following regression model with a discrete hidden logistic process $\bz=(z_1,\ldots,z_n)$, where $z_i\in\{1,\ldots,K \}$:
\begin{equation}
x_i=  \bbeta^T_{z_i}\bsr_{i} + \varepsilon_{i} \quad  ; \quad  i=1,\ldots,n\quad
\label{eq.regression model}
\end{equation}
 In this model, $\bbeta_{z_i}$ is the $(p+1)$-dimensional coefficients vector of a $p$ degree polynomial, $\bsr_{i}=(1,t_i,\ldots,(t_i)^p)^T$ is the time dependant $(p+1)$-dimensional covariate vector associated to the parameter $\bbeta_{z_i}$ and the $\varepsilon_{i}$ are independent random variables distributed according to a Gaussian distribution with zero mean and variance $\sigma^2_{z_{i}}$.

\subsection{The hidden logistic process}
\label{ssec: process}

This section defines the probability distribution of the process $\bz=(z_1,\ldots,z_n)$ that allows the switching from one regression model to another. The proposed hidden logistic process supposes that the variables $z_i$, given the vector $\bst=(t_1,\ldots,t_n)$, are generated independently according to the multinomial distribution {\small$\mathcal{M}(1,\pi_{i1}(\bw),\ldots,\pi_{iK}(\bw))$}, where
\begin{equation}\pi_{ik}(\bw)= p(z_i=k;\bw)=\frac{\exp{(\bsw_k^T\bsv_{i})}}{\sum_{\ell=1}^K\exp{(\bsw_{\ell}^T \bsv_{i})}}\end{equation} is the logistic transformation of a linear function of the time-dependant covariate $\bsv_i=(1,t_i,\ldots,(t_i)^q)^T$, $\bsw_{k}=(\bsw_{k0},\ldots,\bsw_{kq})^T$ is the $(q+1)$-dimensional coefficients vector associated to the covariate $\bsv_i$ and $\bw = (\bsw_1,\ldots,\bsw_K)$. Thus, given the vector $\bst=(t_1,\ldots,t_n)$, the distribution of $\bz$ can be written as:
\begin{equation}
p(\bz;\bw)=\prod_{i=1}^n \prod_{k=1}^K \left(\frac{\exp{(\bsw_{k}^T\bsv_{i})}}{\sum_{\ell=1}^K\exp{(\bsw_{\ell}^T \bsv_{i})}}\right )^{z_{ik}} \enspace ,\\
\end{equation}
where $z_{ik} = 1$ if $z_i=k$ i.e when $x_i$ is generated by the $k^{th}$ regression model, and $0$ otherwise.

\begin{figure}[!h]
\begin{tabular}{ccc}
\includegraphics[height = 3 cm,width=3.7cm]{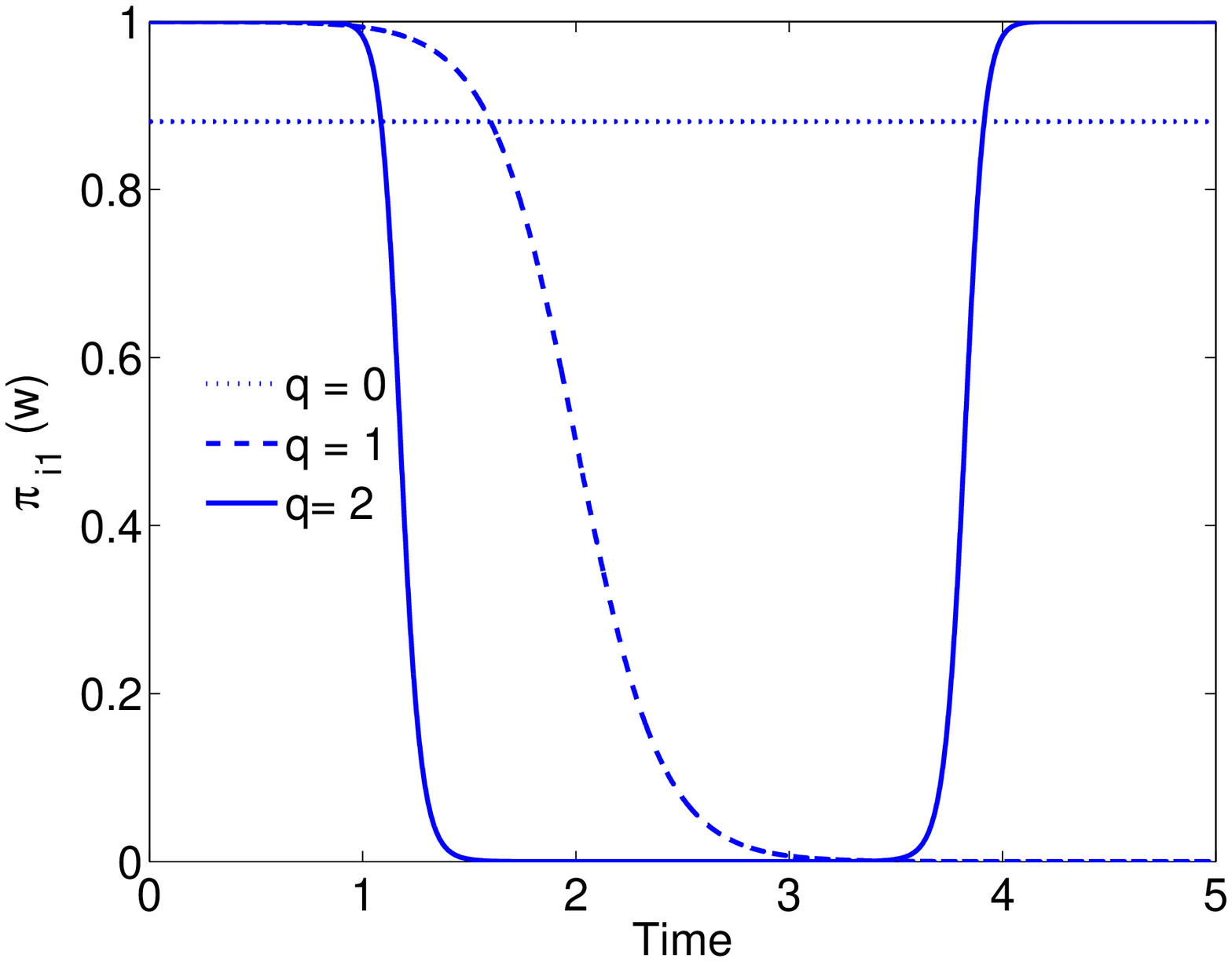} &
\includegraphics[height = 3 cm,width=3.7cm]{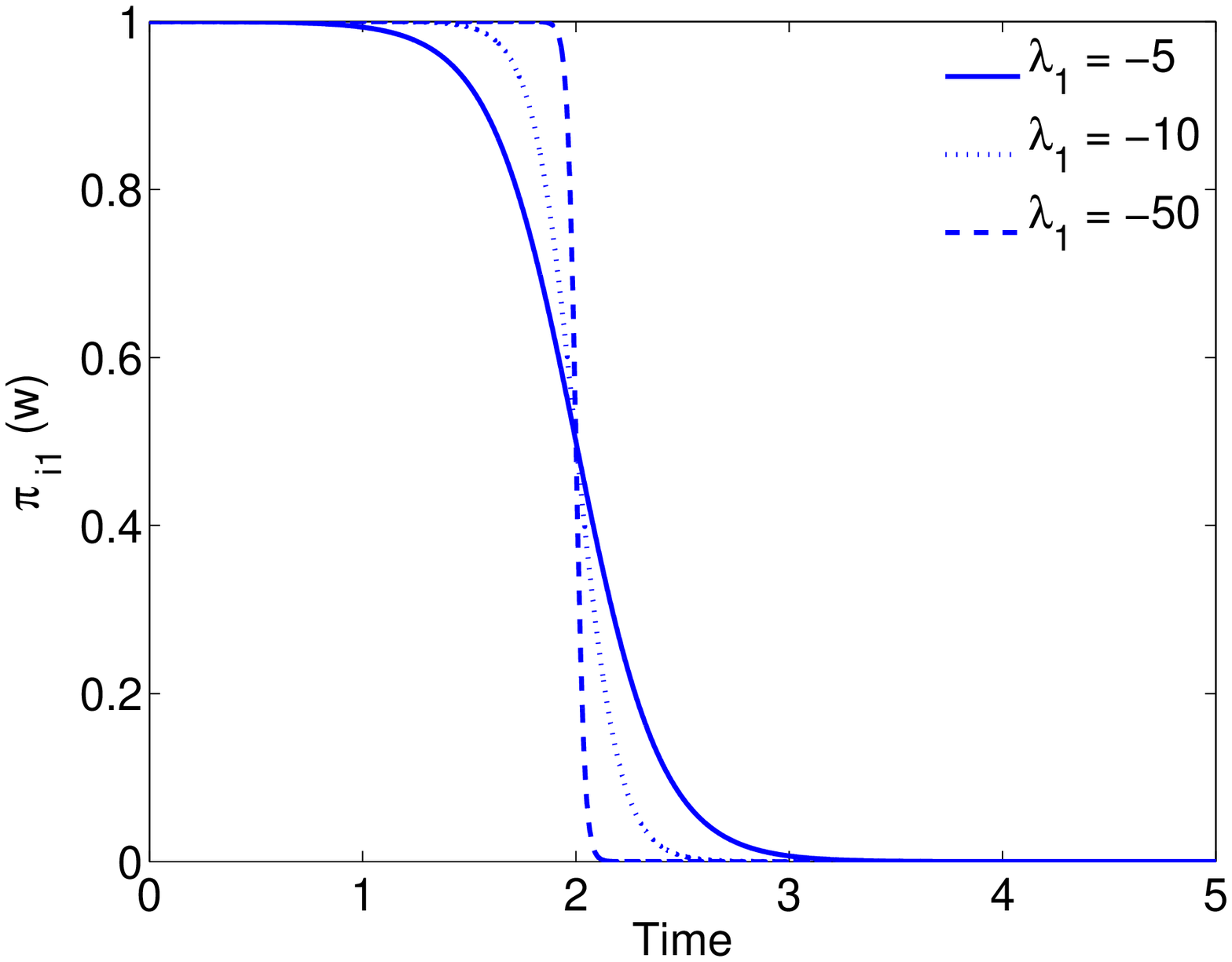} &
\includegraphics[height = 3 cm,width=3.7cm]{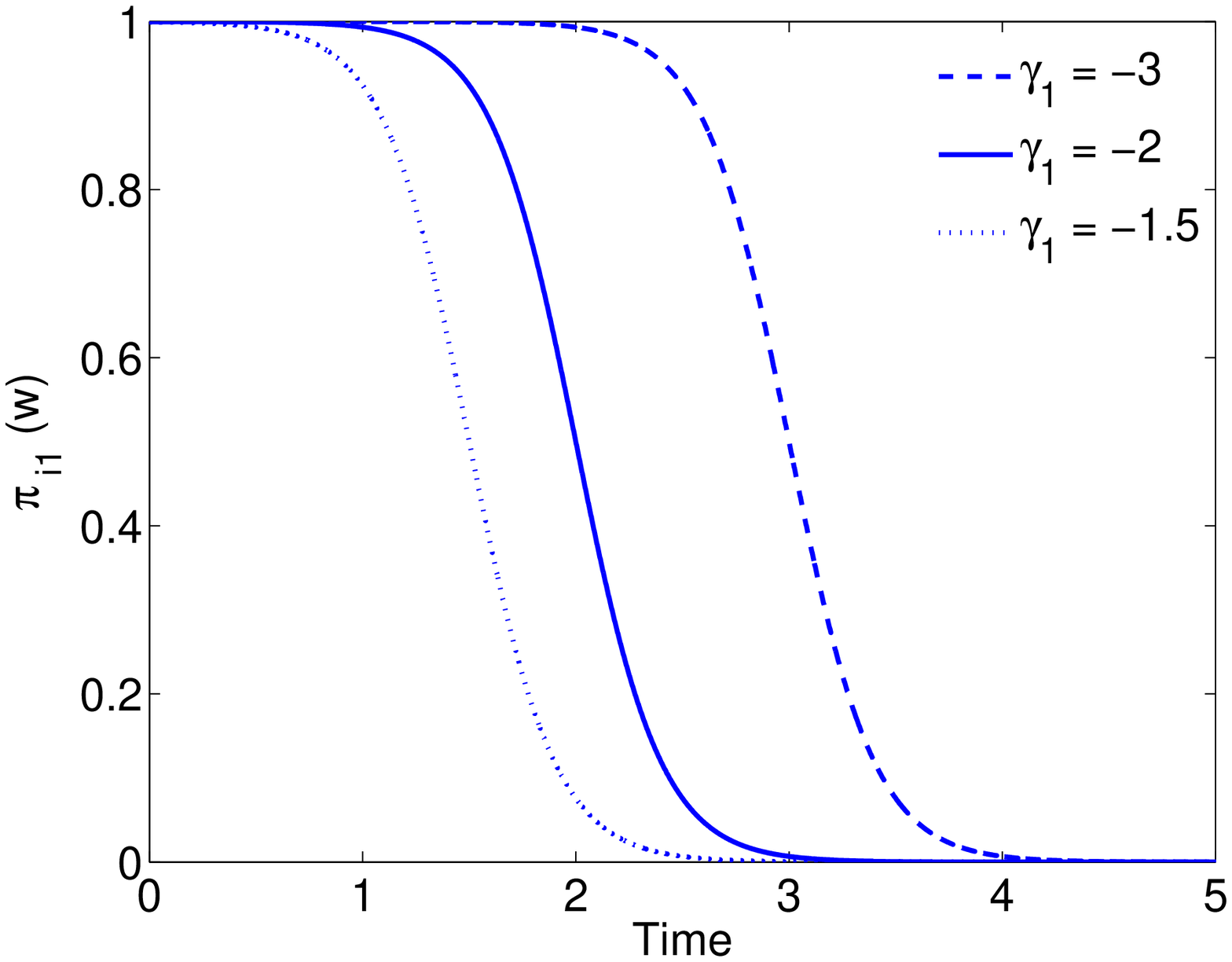}\\
\end{tabular}
\caption{Variation of $\pi_{i1}(\bw)$ over time for (left) different values of the dimension $q$ of $\bsw_k$, (middle) different values of $\lambda_k = \bsw_{k1}$ and (right) different values of $\gamma_k = \frac{\bsw_{k0}}{\bsw_{k1}}$ with $q=1$.}
\label{logistic_function_k=2_p=012}
\end{figure}

The pertinence of the logistic transformation in terms of flexibility of transition can be illustrated through simple examples with $K=2$ components. As it can be shown in figure \ref{logistic_function_k=2_p=012} (left), the dimension $q$ of $\bsw_k$ controls the number of changes in the temporal variation of $\pi_{ik}(\bw)$. More particularly, if the goal is to segment the signals into convex homogenous parts, the dimension $q$ of $\bsw_k$ must be set to $1$. The quality of transitions and the change time point are controlled by the components values of the vector $\bsw_k$ (see figures \ref{logistic_function_k=2_p=012} (middle) and (right)).

\section{Parameter estimation}
\label{sec: parameter estimation}

From the model given by equation (\ref{eq.regression model}), it can be proved that the random variable $x_{i}$ is distributed according to the normal mixture density
\begin{equation}
p(x_{i};\btheta)=\sum_{k=1}^K\pi_{ik}(\bw)\mathcal{N}\big(x_{i};\bbeta^T_k\bsr_{i},\sigma^2_k\big) \enspace,
\label{melange}
\end{equation}
where $\btheta=(\bsw_{1},\ldots,\bsw_{K},\bbeta_1,\ldots,\bbeta_K,\sigma^2_1,\ldots,\sigma^2_K)$
is the parameter vector to be estimated. The parameter $\btheta$ is estimated by the maximum likelihood method. As in the classic regression models we assume that, given $\bst =(t_1,\ldots,t_n)$,  the $\varepsilon_i$ are independent. This also involves the independence of $x_i$ $(i=1,\ldots,n)$. The log-likelihood of $\btheta$ is then written as:
\begin{equation}
L(\btheta;\bx)=\log \prod_{i=1}^np(x_i;\btheta)=\sum_{i=1}^{n}\log\sum_{k=1}^K \pi_{ik}(\bw)\mathcal{N}\big(x_{i};\bbeta^T_k\bsr_i,\sigma^2_k\big)
\end{equation}
Since the direct maximization of this likelihood is not straightforward, we use the Expectation Maximization (EM) algorithm \cite{dlr}\cite{mclachlan EM} to perform the maximization.

\subsection{The dedicated EM algorithm}
\label{ssec. EM algortihm}
The proposed EM algorithm starts from an initial parameter $\btheta^{(0)}$ and alternates the two following steps until convergence:
\paragraph{\textbf{E Step (Expectation):}}
This step consists of computing the expectation of the complete log-likelihood $\log p(\bx,\bz;\btheta)$, given the observations and the current value $\btheta^{(m)}$ of the parameter $\btheta$ ($m$ being the current iteration):
\begin{eqnarray}
Q(\btheta,\btheta^{(m)})&=& E[\log p(\bx,\bz;\btheta)|\bx;\btheta^{(m)}]\nonumber\\
&=&\sum_{i=1}^{n}\sum_{k=1}^K t^{(m)}_{ik}\log (\pi_{ik}(\bw)\mathcal{N}(x_{i};\bbeta^T_k\bsr_{i},\sigma^2_k)) \enspace,
\end{eqnarray}
where
$ t^{(m)}_{ik}=p(z_{ik}=1|x_i;\btheta^{(m)})=\frac{\pi_{ik}(\bw^{(m)})\mathcal{N}(x_{i};\bbeta^{T(m)}_k\bsr_{i},\sigma^{2(m)}_k)}
{\sum_{\ell=1}^K\pi_{i \ell}(\bw^{(m)})\mathcal{N}(x_{i};\bbeta^{T(m)}_{\ell}\bsr_{i},\sigma^{2(m)}_{\ell})}
$
is the posterior probability that $x_i$ originates from the $k^{th}$ regression model. As shown in the expression of $Q$, this step simply requires the computation of $t^{(m)}_{ik}$.
\paragraph{\textbf{M step (Maximization):}}
In this step, the value of the parameter $\btheta$ is updated by computing the parameter $\btheta^{(m+1)}$ maximizing the expectation $Q$
with respect to $\btheta$. The maximization of $Q$ can be performed by separately maximizing
\begin{equation} \small{{Q_1(\bw)=\sum_{i=1}^{n}\sum_{k=1}^K t^{(m)}_{ik}\log \pi_{ik}(\bw)}  ~~\mbox{and}~~ \small{Q_2(\bbeta_k,\sigma^2_k)=\sum_{i=1}^{n} t^{(m)}_{ik}\log \mathcal{N}(x_{i};\bbeta^T_k\bsr_{i},\sigma^2_k)}} 
\end{equation}

The maximization of $Q_1$ with respect to $\bw$ is a multinomial logistic regression problem weighted by the $t^{(m)}_{ik}$. We use a multi-class Iterative Reweighted Least Squares (IRLS) algorithm \cite{irls}\cite{krishnapuram}\cite{chen} to solve it. Maximizing  $Q_2$ with respect to $(\bbeta_k,\sigma^2_k)$ consists of analytically solving a weighted least-squares problem.
\!
\subsection{Denoising and segmenting a signal}
\label{sse: estimation of the denoised signal}
\!
In addition to providing a signal parametrization, the proposed approach can be used to denoise and segment signals.
The denoised signal can be approximated by the expectation
\begin{equation}
E(x_i;\hat{\btheta}) = \int_{\IR}x_i p(x_i;\hat{\btheta})dx_i
			= \sum_{k=1}^{K} \pi_{ik}(\hat{\bw})\hat{\bbeta}^T_k \bsr_{i}\enspace,\quad  \forall i=1,\ldots,n
\label{eq. signal expectation}
\end{equation}
where $\hat{\btheta} = (\hat{\bw},\hat{\bbeta}_1,\ldots,\hat{\bbeta}_K,\hat{\sigma}^2_1,\ldots,\hat{\sigma}^2_K)$ is the parameters vector obtained at the convergence of the algorithm. On the other hand, a signal segmentation  can also be deduced by computing the estimated label $\hat{z_i}$ of $x_i$: $\hat{z_i} = \arg \max \limits_{\substack {1\leq k\leq K}} \pi_{ik}(\widehat{\bw})$.

\section{Experiments}
\label{sec: experiments}

This section is devoted to the evaluation of the proposed algorithm using simulated and real data sets. Two evaluation criteria are used in the simulations: the misclassification rate between the simulated partition and the estimated partition and the euclidian distance between the denoised simulated signal and the estimated denoised signal normalized by the sample size $n$. 
 The proposed approach is compared to  the piecewise regression approach \cite{McGee} .
\subsection{Simulated signals}
\begin{figure}[!h]
\centering
\includegraphics[height = 3.3 cm, width=4.5cm]{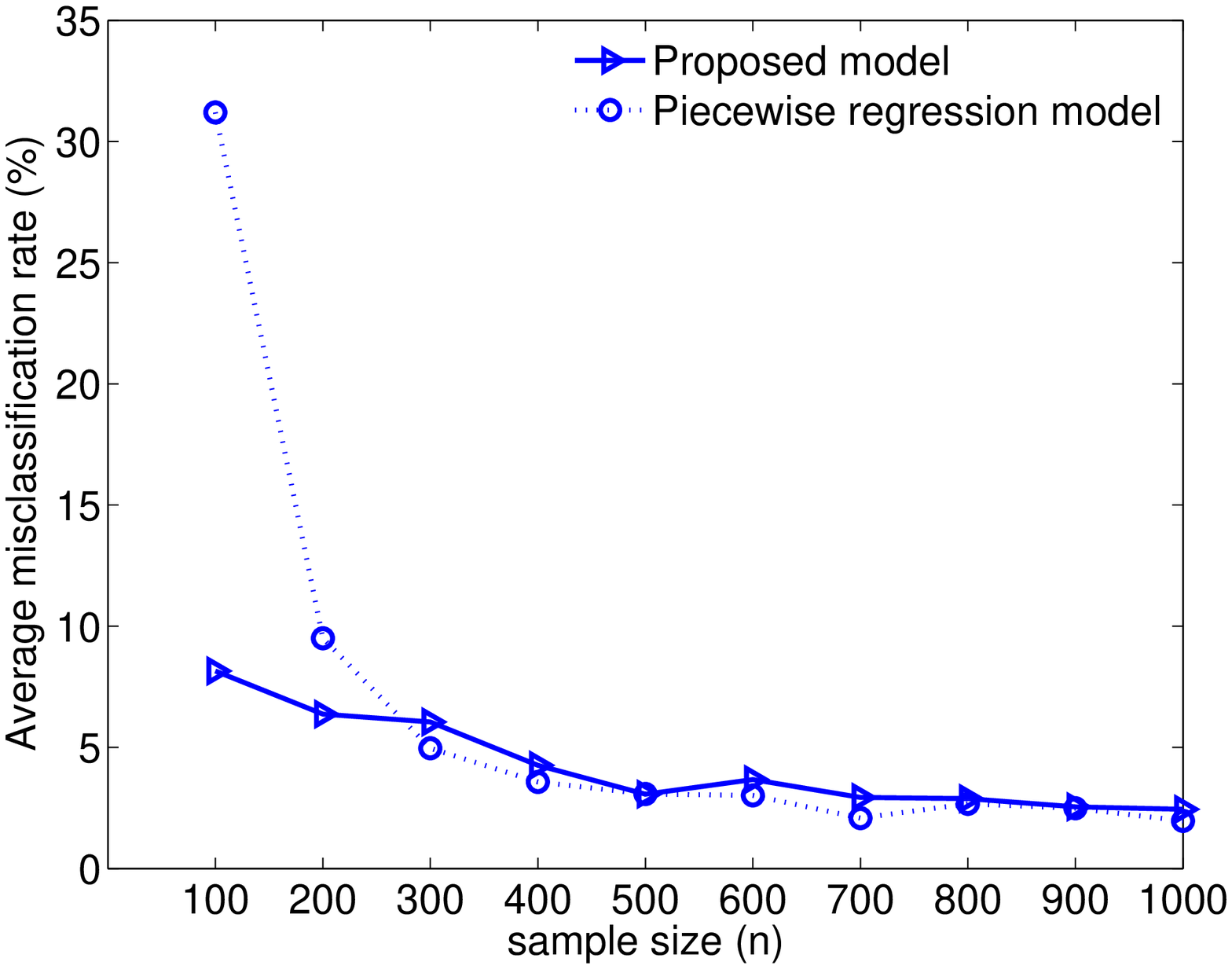}
\hspace{2cm}
\includegraphics[height = 3.3 cm, width=4.5cm]{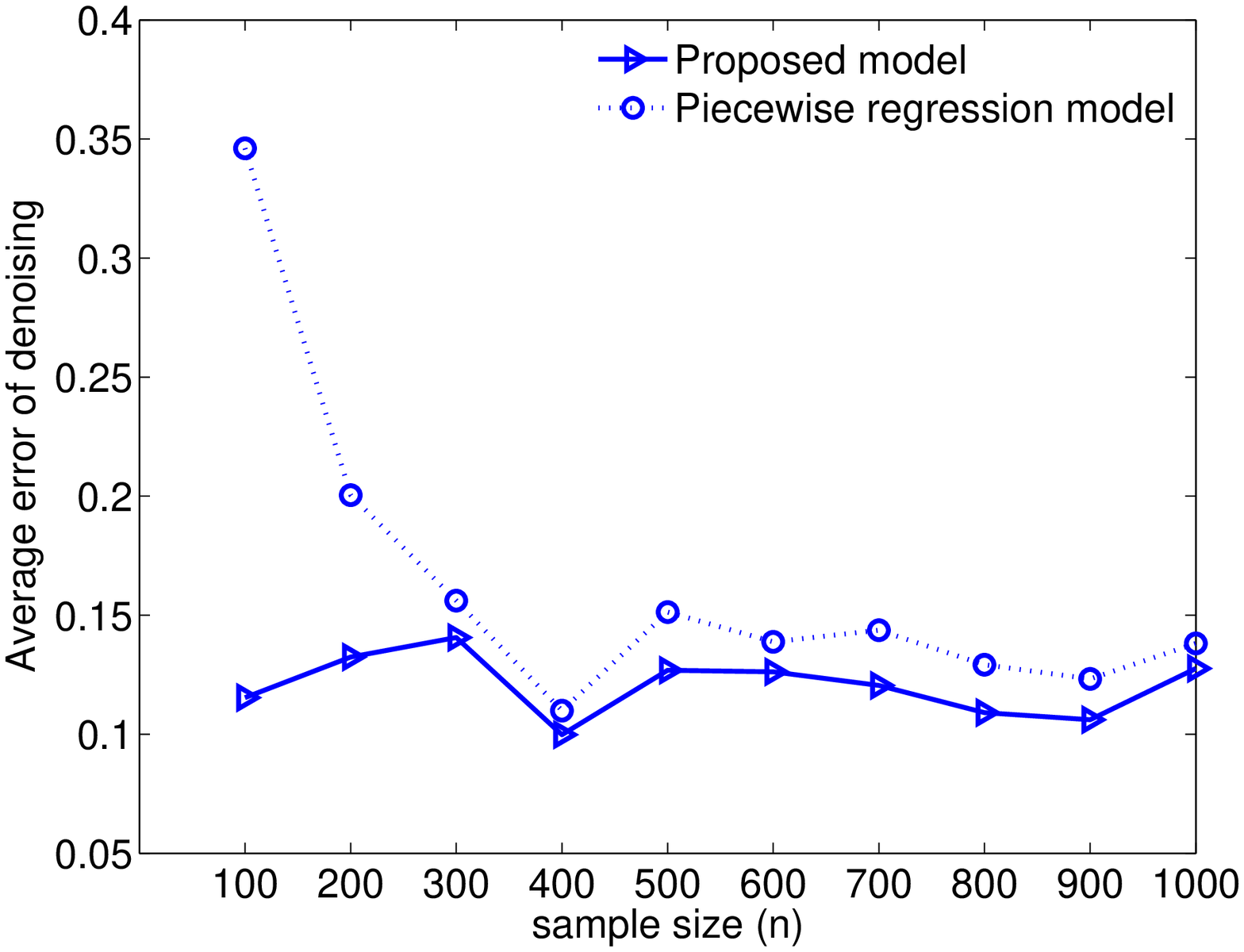} 
\caption{Average values of misclassification rate (left) and error of denoising (right) in relation to the sample size $n$ obtained with the proposed approach (triangle) and the piecewise regression approach (circle).}
\label{fig.error rates}
\end{figure}

Each signal is generated according to the regression model  with a hidden logistic process defined by eq (\ref{eq.regression model}). The number of states of the hidden variable is fixed to $K=3$ and the order of regression is set to $p=2$. The order of the logistic regression is fixed to $q=1$ what guarantees a segmentation into convex intervals. We consider that all signals are observed over $5$ seconds. For each size $n$ we generate 20 samples. The values of assessment criteria are averaged over the 20 samples. Figure \ref{fig.error rates} (left) shows the misclassification rate obtained by the two approaches in relation to the sample size $n$. It can be observed that the proposed approach is more stable for a few number of observations. Figure \ref{fig.error rates} (right) shows the results obtained by the two approaches in terms of signal denoising. It can be observed that the proposed approach provides a more accurate denoising of the signal compared to the piecewise regression approach. For the proposed model, the optimal values of $(K,p)$ has also been estimated by computing the Bayesian Information Criterion (BIC) \cite{BIC criterion} for $k$ varying from $2$ to $K_{max}=8$ and $p$ varying from $0$ to $p_{max}=6$. The simulated model, corresponding to $K=3$ and $p=2$, has been chosen with the maximum percentage of $85\%$.

\subsection{Real signals}
\label{ssec: exp on real data}

\begin{figure}[!h]
\centering
\includegraphics[height = 3.3 cm, width=3.9cm]{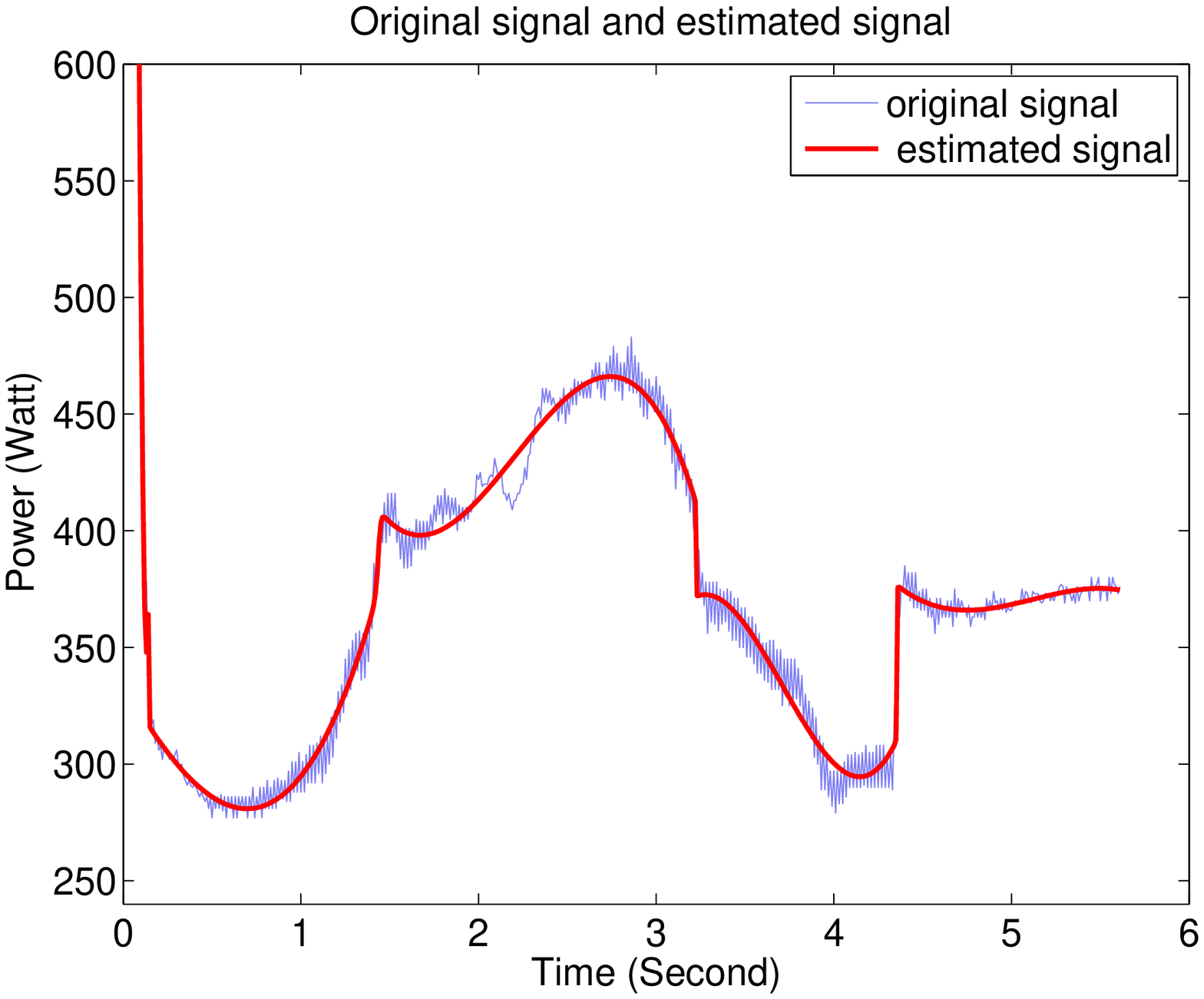}
\includegraphics[height = 3.3 cm, width=3.9cm]{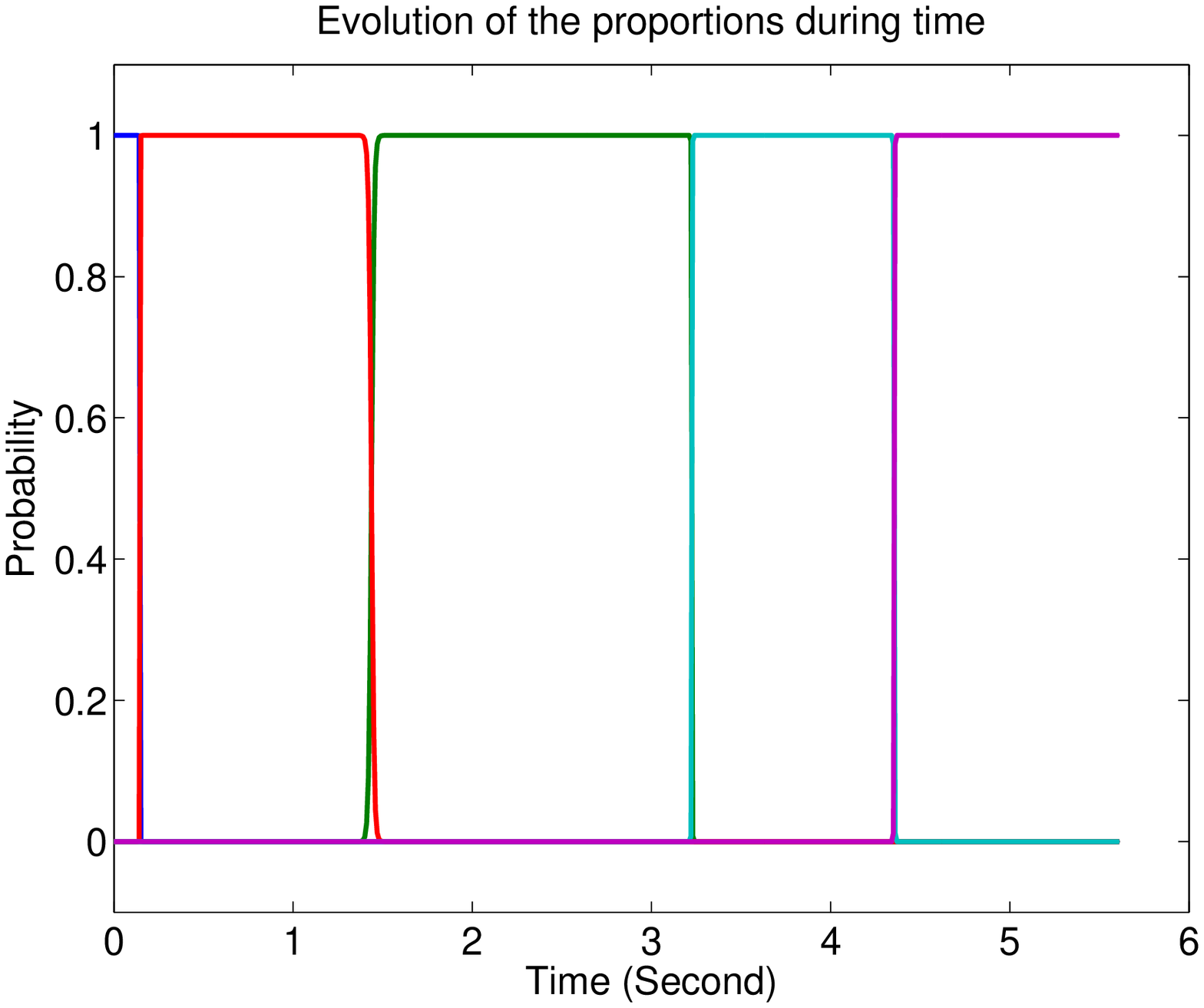}
\includegraphics[height = 3.3 cm, width=3.9cm]{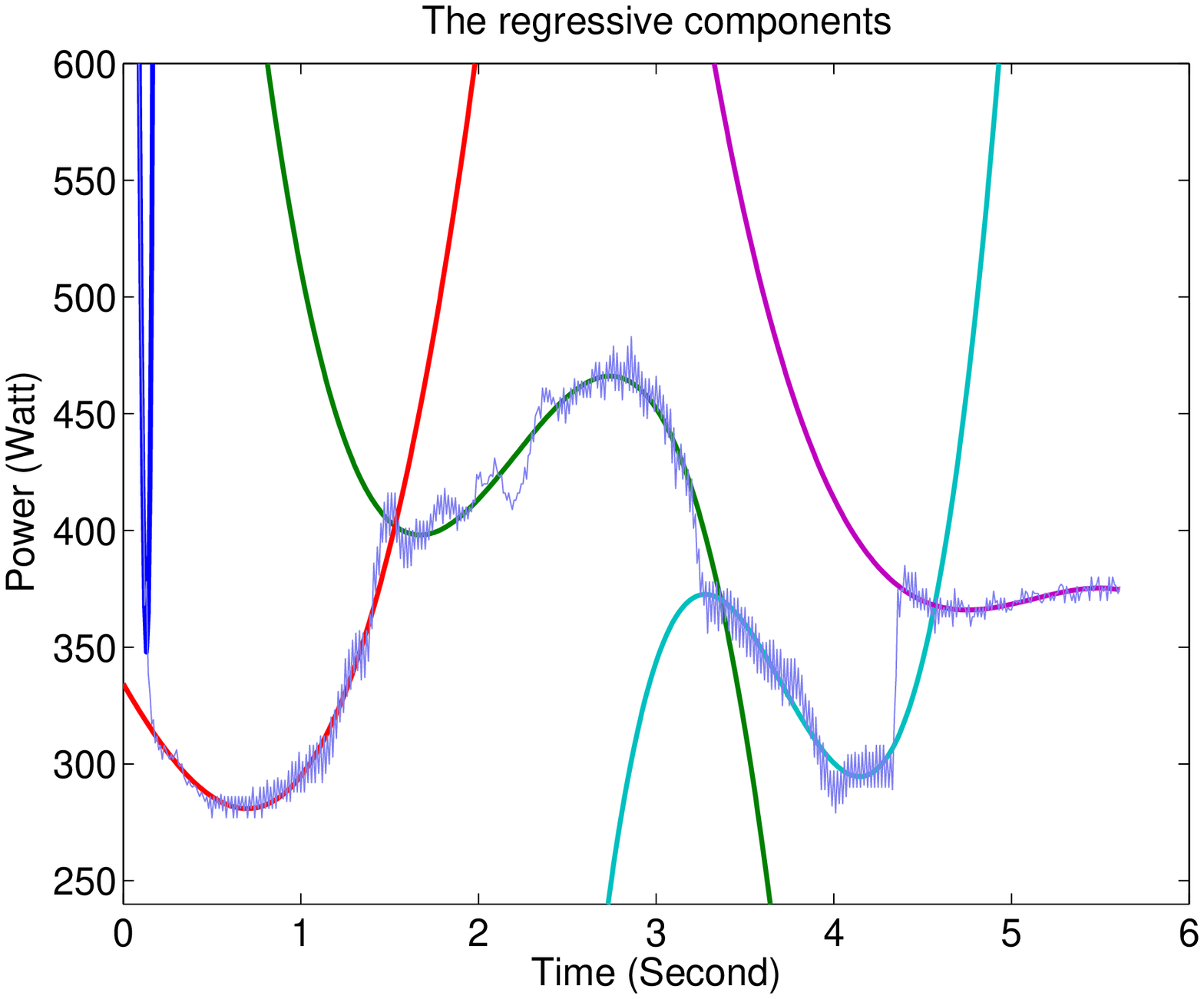}
\caption{Results obtained for a signal with defect}
\label{resultat_signal_aig_2}
\end{figure}

This section presents the results obtained by the proposed model for signals of switch points operations. One situation corresponding to a signal with a critical defect is presented. The number of the regressive components is chosen in accordance with the number of the electromechanical phases of a switch points operation ($K = 5$). The value of $q$ has been set to $1$, what guarantees a segmentation into convex intervals, and the degree of the polynomial regression has been set to $3$ which is adapted to the different regimes in the signals. Figure \ref{resultat_signal_aig_2} (left) shows the original signal and the denoised signal (given by equation (\ref{eq. signal expectation})). Figure \ref{resultat_signal_aig_2} (middle) shows the variation of the proportions $\pi_{ik}$ over time. It can be observed that these probabilities are very closed to $1$ when the $k^{th}$ regressive model seems to be the most faithful to the original signal. The five regressive components involved in the signal are shown in figure \ref{resultat_signal_aig_2} (right).

\section{Conclusion}
\label{sec: conclusion}
In this paper a new approach for signals parametrization, in the context of the railway switch mechanism monitoring, has been proposed. This approach is based on a regression model incorporating a discrete hidden logistic process. The logistic probability function, used for the hidden variables, allows for smooth or abrupt switchings between polynomial regressive components over time. In addition to signals parametrization, an accurate denoising and segmentation of signals can be derived from the proposed model. 


\begin{footnotesize}
%


\end{footnotesize}



\begin{thebibliography}{99}
\bibitem{dlr} A.~P.~Dempster, N.~M.~Laird and D.~B.~Rubin, Maximum likelihood from incomplete data via the EM algorithm, \emph{Journal of the Royal Statistical Society,} B, 39(1): 1-38, 1977.


\bibitem{krishnapuram} B. Krishnapuram, L. Carin, M.A.T. Figueiredo and A.J. Hartemink, Sparse multinomial logistic regression: fast algorithms and generalization bounds, \emph{IEEE Transactions on Pattern Analysis and Machine Intelligence,} 27(6): 957-968, June 2005.

\bibitem{mclachlan EM}  G. J. McLachlan and T. Krishnan, \emph{The EM algorithm and extensions,} Wiley series in probability and statistics, New York, 1997.

\bibitem{BIC criterion} G. Schwarz, Estimating the dimension of a model, \emph{Annals of Statistics,} 6: 461-464, 1978.

\bibitem{chen} K. Chen, L. Xu and H. Chi, Improved learning algorithms for Mixture of Experts in multiclass classification, \emph{IEEE Transactions on Neural Networks,} 12(9): 1229-1252, November 1999.



\bibitem{fridman} M. Fridman, Hidden Markov Model Regression, Technical Report, Institute of mathematics, University of Minnesota, December 1993.

\bibitem{jordan HME} M.~I.~Jordan and R.~A.~Jacobs, Hierarchical Mixtures of Experts and the EM algorithm, \emph{Neural Computation,} 6: 181-214, 1994.

\bibitem{irls} P.~Green. Iteratively Reweighted Least Squares for Maximum Likelihood Estimation, and some robust and resistant alternatives, \emph{Journal of the Royal Statistical Society,} B, 46(2): 149-192, 1984.


\bibitem{quandt} R. E. Quandt and and J. B. Ramsey, Estimating mixtures of normal distributions and switching regressions, \emph{Journal of the American Statistical Association,} 73(364): 730-752, 1978.

\bibitem{waterhouse} S. R. Waterhouse, \emph{Classification and regression using Mixtures of Experts,} PhD thesis, Department of Engineering, Cambridge University, 1997.
\bibitem{McGee} V. E. McGee and  W. T. Carleton, Piecewise regression, \emph{Journal of the American Statistical Association}, 65, 1109-1124, 1970.

\end{thebibliography}
\end{document}